\documentclass[12pt]{article}
\usepackage[dvips]{graphicx}
\renewcommand{\baselinestretch}{1.20}
\begin{document}

%
\begin{flushright}
OU-HET-657, \, March, 2010 \ \ \\
\end{flushright}
\vspace{0mm}
\begin{center}
\large{Flavor asymmetry of the polarized sea-quark distributions
in the proton}
\end{center}
\vspace{0mm}
\begin{center}
M.~Wakamatsu\footnote{Email \ : \ wakamatu@phys.sci.osaka-u.ac.jp}
\end{center}
\vspace{-4mm}
\begin{center}
Department of Physics, Faculty of Science, \\
Osaka University, \\
Toyonaka, Osaka 560-0043, JAPAN
\end{center}


\vspace{6mm}
\begin{center}
\small{{\bf Abstract}}
\end{center}
\vspace{-2mm}
\begin{center}
\begin{minipage}{15.5cm}
\renewcommand{\baselinestretch}{1.0}
\small
The recent global analysis of helicity parton distributions,
which takes into account available data from inclusive and
semi-inclusive polarized deep inelastic scattering, as well
as from polarized proton-proton scattering at RHIC, appears to
offer the first strong evidence that polarized sea-quark
distributions are flavor asymmetric, i.e.
$\Delta \bar{u}(x) \ne \Delta \bar{d}(x)$.
We point out that the flavor symmetry breaking pattern indicated by
their analysis, i.e. $\Delta \bar{u}(x) > 0$ and
$\Delta \bar{d}(x) < 0$ with the magnitude correlation
$|\Delta \bar{u}(x)| < |\Delta \bar{d}(x)|$, is just consistent
with our theoretical predictions given several
years ago on the basis of the chiral quark soliton model.
We also address ourselves to understanding the physics behind this
observation.

\normalsize
\end{minipage}
\end{center}
\renewcommand{\baselinestretch}{2.0}

\vspace{8mm}
Undoubtedly, the famous NMC measurement \cite{NMC91}, which has
established the flavor asymmetry of unpolarized sea-quark
distributions, is thought to be one of the most noticeable
achievements in the recent studies of nucleon structure functions.
The reason is that it gave the first clear evidence for
manifestation of nonperturbative chiral dynamics of QCD in
high-energy deep-inelastic scattering observables, which was
not taken very seriously before this milestone discovery. 
The NMC observation, i.e. the excess of $\bar{d}$-sea over the
$\bar{u}$-sea in the proton, is known to be explained by a variety
of models at least qualitatively.
(See \cite{Kumano98},\cite{GP01}, for review.)  
They are the meson cloud convolution model including its
variants \cite{Sullivan72}\nocite{Thomas83}\nocite{HM90}
\nocite{Kumano91}\nocite{MT91}-\cite{HSS96},
the chiral quark soliton model (CQSM) \cite{Waka91}\nocite{Waka92}
\nocite{WT98}-\cite{PPGWW99}, as well as
several other models with more phenomenological nature like
the statistical parton model \cite{Bhalerao01},\cite{BSB02} and
the explanation based on the Pauli exclusion
principle \cite{GR00}.

A natural next question is then whether the polarized antiquark
sea in the nucleon is also flavor asymmetric or not.
Somewhat embarrassingly, existing theoretical
answers for this question is fairly dispersed, in remarkable
contrast to the unpolarized case. 
Among others. worthy of special mention is a big difference
between the prediction of the meson cloud model \cite{FS98}
\nocite{BK99}\nocite{CS01}\nocite{KM02}\nocite{FSW03}-\cite{CS03}
and that of the CQSM  \cite{DPPPW9697}
\nocite{WGR9697}\nocite{WK99}\nocite{WW00}\nocite{DGPW00}
-\cite{Waka03AB}. The CQSM predicts large flavor asymmetry also
for the longitudinally polarized sea-quark distributions, i.e.
$\Delta \bar{u}(x) - \Delta \bar{d}(x) > 0$, whereas the prediction
of the meson cloud model for the same quantity is fairly small or
slightly positive. (Actually, the meson cloud model contains a lot
of parameters and its theoretical predictions are fairly dispersed
depending on how many meson-baryon intermediate channels 
are included in the calculation \cite{FSW03}.
Here, we are supposing the
prediction of the most elaborate recent calculation
by Cao and Signal within the framework of the meson cloud
model \cite{CS03}.)
Although it is usually believed that the CQSM yield similar
results to those obtained in the meson cloud model or the cloudy
bag model as one of such models constructed so as to effectively
incorporate the chiral dynamics of Nambu-Goldstone excitations
surrounding the nucleon core, it is not necessarily true.
Significant differences, if they exist, appears to originate
from a unique dynamical ansatz of the CQSM, i.e. the
rotating hedgehog. It has been long claimed that this unique
feature of the CQSM enables us to explain the celebrated EMC
observation, i.e. very small quark spin fraction of the nucleon,
quite naturally without introducing any fine
tuning \cite{WY91},\cite{WW00L}.
The recently found big difference between the prediction of the
CQSM and that of the refined cloudy bag model for the isovector
combination of the quark orbital angular momenta
$L^u - L^d$ \cite{Waka09},\cite{Thomas09}
also appears to be connected with the nontrivial spin-isospin
correlation between quark fields embedded in the hedgehog
ansatz \cite{Waka09},\cite{WT05},\cite{WN0608}.
Furthermore, the same correlation between spin and isospin is
likely to be the cause of the strong  correlation existing between
the flavor asymmetries of the unpolarized and polarized sea-quark
distributions predicted by the CQSM, which dictates that
both of $\bar{u}(x) - \bar{d}(x)$ and
$\Delta \bar{u}(x) - \Delta \bar{d}(x)$ are sizably large
in magnitude \cite{WW00},\cite{DGPW00}. 
    
In view of the interesting sensitivity of the flavor asymmetry
of the polarized sea-quark distributions to theoretical models,
it is of great interest to get direct experimental information on it.
Since the separation of the quark and antiquark distributions
cannot be done solely from the inclusive measurements, additional
information from semi-inclusive measurements is crucial for
this separation. 
The first systematic challenge for aiming at extracting the polarized
sea-quark distributions were carried out by the HERMES
Collaboration \cite{HERMES0504}.
From semi-inclusive scattering measurements where the final pions and
kaons are measured, they extracted the polarized antiquark
distributions, thereby concluding that the polarization of each flavor,
i.e. $\Delta \bar{u}(x), \Delta \bar{d}(x), \Delta \bar{s}(x)$,
is very small, and compatible with zero, which seems to be consistent
with the prediction of the meson cloud model \cite{CS03}.
However, in view of the fact that the mechanism of semi-inclusive
scatterings is understood less reliably than that of
the inclusive scatterings, and that we have much more precise inclusive
data than the semi-inclusive data,
it is desirable to perform systematic global analysis, which takes
account of all the available information.
Such an analysis has recently been done by de Florian et al.
\cite{DSSV09}
Their analysis was performed fully at the next-to-leading order
of perturbative QCD, by taking account of available data from
inclusive and semi-inclusive scatterings, as well as from polarized
proton-proton scatterings at RHIC. Very interestingly, the result
of their analysis appears to offer the first strong evidence in
favor of the flavor asymmetry of the polarized sea-quark distributions,
i.e. $\Delta \bar{u}(x) - \Delta \bar{d}(x) > 0$.
Particularly noteworthy here is the observed pattern of flavor symmetry
violation in the polarized sea. Their results indicates that
$\Delta \bar{u}(x) > 0$ and $\Delta \bar{d}(x) < 0$ with the
interesting magnitude correlation
$|\Delta \bar{u}(x)| < |\Delta \bar{d}(x)|$.

Now, the purpose of the present paper is to point out that
the observed pattern of flavor symmetry violation in the polarized
sea-quark distribution is just consistent with the parameter-free
predictions of the CQSM, which we gave several years ago.
We also try to clarify the background physics leading to the
observed symmetry breaking pattern of the polarized sea-quark
distributions.
We shall also make a short remark on their results for the
polarized strange-quark distributions in the nucleon from our
own viewpoint.

Before showing a comparison of the predictions of the CQSM
with the results of the new DSSV analysis, several comments on the
model are in order.
We have two versions of the CQSM. One is the flavor SU(2)
version \cite{DPP88},\cite{WY91},
and the other is the flavor SU(3)
version \cite{WAR92},\cite{BDGPPP93}. The basic parameter
common in both models is the dynamically generated quark mass $M$,
which is already fixed to be $M \simeq 375 \,\mbox{MeV}$ from low
energy phenomenology or from the instanton picture of the QCD vacuum,
which affords a theoretical foundation of the model \cite{DPP88}.
The predictions of the SU(2) CQSM for various parton distributions
at the model scale is therefore parameter free.
The flavor SU(3) version of the CQSM model contains an additional
parameter, i.e. the effective mass difference $\Delta m_s$ between
the strange and up-down quarks. We fixed this parameter to be
$100 \,\mbox{MeV}$ such that the model reproduces the general
behavior of the unpolarized strange quark distributions at the
high energy scale, $Q^2 = 4 \,\mbox{GeV}^2$. (See \cite{Waka03AB},
for more detail.) To make a comparison with high energy
deep-inelastic-scattering observables, we take the predictions of
the CQSM as initial scale distributions at the low energy model
scale. The scale dependencies of the distribution functions
are taken into account by using the standard evolution equation at
the next-to-leading order. The starting energy of this evolution
is taken to be $Q^2 = 0.30 \,\mbox{GeV}^2$, basically following
the strategy of the PDF fits by Gl\"{u}ck, Reya and
Vogt \cite{GRV95},\cite{GRSV96}. 

\begin{figure}[htb]
\begin{center}
  \includegraphics[height=.40\textheight]{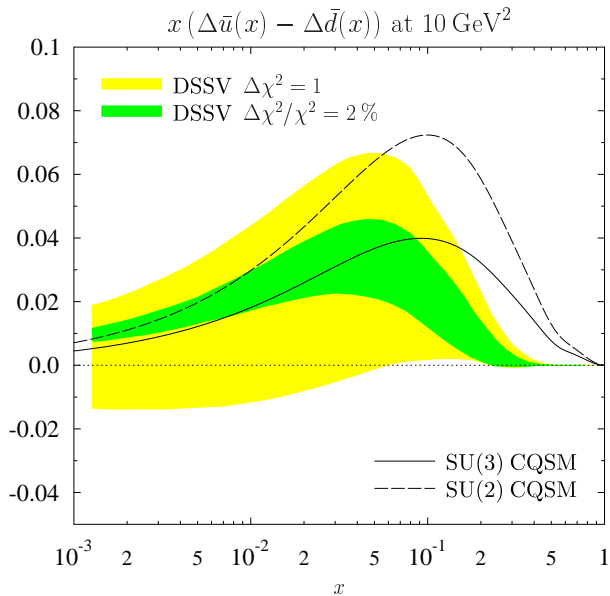}
  \caption{\baselineskip16pt The predictions of the $SU(2)$ and
  $SU(3)$ CQSM for the distribution difference
  $x \,(\Delta \bar{u}(x) - \Delta \bar{d}(x))$
  at $Q^2 = 10 \,\mbox{GeV}^2$, in comparison with the new DSSV
  global fit with the uncertainty bands for $\Delta \chi^2 = 1$
  and $\Delta \chi^2 / \chi^2 = 2 \,\%$.}%
\label{fig1:xDelubmdb}
\end{center}
\end{figure}

Now in Fig.1, we show the results of the new DSSV global fit for the
isovector distribution $x \,(\Delta \bar{u}(x) - \Delta \bar{d}(x))$
in comparison with the theoretical predictions of the CQSM.
Here, the solid and dash-dotted curves are respectively the
predictions of the flavor SU(2) and SU(3) versions of the CQSM.
As pointed out in \cite{Waka03AB}, the flavor asymmetry of the polarized
sea-quark distributions is fairly sensitive to the difference
of the two versions of the model. (This is not the case for the
unpolarized sea-quark distributions. That is, the difference
between the predictions of the two models for the distribution
$x \,(\bar{u}(x) - \bar{d}(x))$ is fairly small \cite{Waka03AB}.)
One sees that, with high confidence level, the new DSSV fit
shows a strong evidence in favor of the flavor symmetry violation for
the polarized sea-quark distributions. Interestingly, the magnitude of
this flavor symmetry violation is fairly close to the prediction 
of the flavor SU(3) CQSM. It was advocated that
sizably large CQSM prediction for
$\Delta \bar{u}(x) - \Delta \bar{d}(x)$ is consistent with the
large-$N_c$ counting argument \cite{DGPW00}, which dictates that
\begin{equation}
 | \Delta \bar{u}(x) - \Delta \bar{d}(x) | \ \gg \ 
 | \bar{u}(x) - \bar{d}(x) | ,
\end{equation}
since $|\bar{u}(x) - \bar{d}(x) | \,/\,
|\Delta \bar{u}(x) - \Delta \bar{d}(x) |$ is a $1 / N_c$ quantity.
In our realistic world, however, $N_c$ is just three, anyway,
and the actual numerical predictions might not necessarily obey
this general expectation. In fact, in the SU(3) CQSM, we find
that the magnitude of $\Delta \bar{u}(x) - \Delta \bar{d}(x)$ is
slightly smaller than that of $\bar{u}(x) - \bar{d}(x)$, as
shown in Fig.18 of \cite{Waka03AB}. (Incidentally, our prediction for
$x \,(\Delta \bar{u}(x) - \Delta \bar{d}(x))$ within the SU(2)
CQSM is a little smaller than the corresponding prediction of the
Bochum group shown in Fig.7 of \cite{DSSV09}. The reason of this small
discrepancy is not clear, but it may be traced back to the
difference of the used soliton profile or the difference of
the details of the evolution procedure.)

\begin{figure}[bht]
\begin{center}
  \includegraphics[height=.40\textheight]{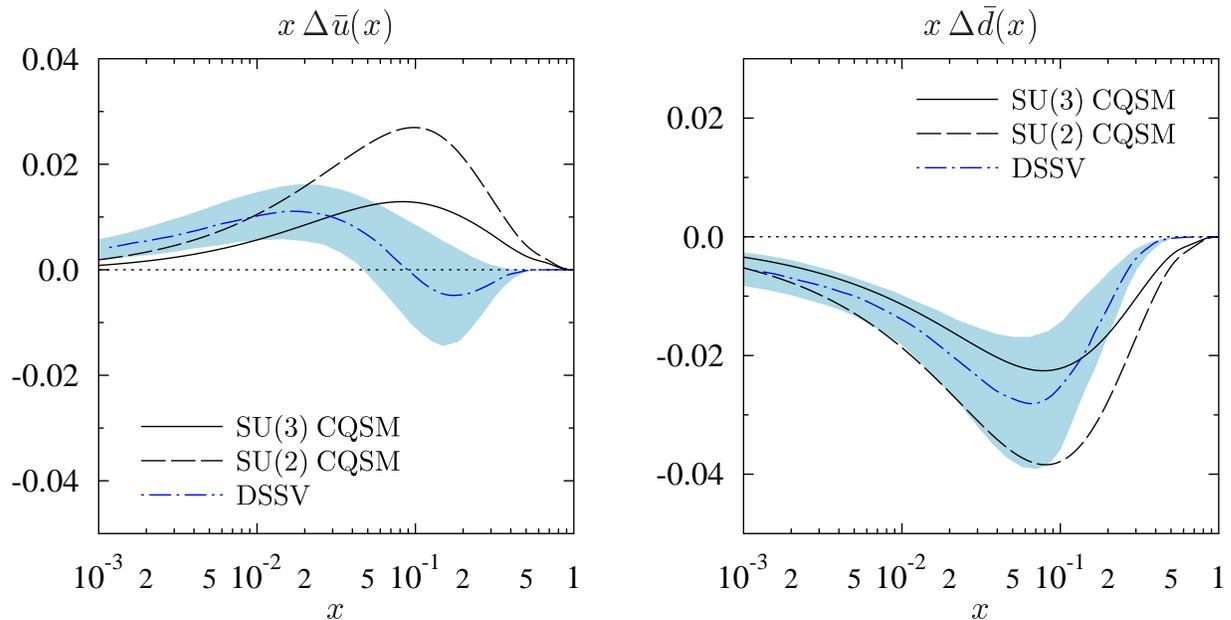}
  \caption{\baselineskip16pt The predictions of the $SU(2)$ and
  $SU(3)$ CQSM for the longitudinally polarized sea-quark distribution
  functions, $x \,\Delta \bar{u}(x)$ and $x \,\Delta \bar{d}(x)$ at
  $Q^2 = 10 \,\mbox{GeV}^2$, in comparison with the
  DSSV global fit.}%
\label{fig2:xDeludbar}
\end{center}
\end{figure}

Also very interesting is the flavor separation of the polarized
sea-quark distributions. Shown in Fig.2 are the results of the DSSV
analysis in comparison with the predictions of the CQSM.
Again, the predictions of the SU(3) CQSM appears to be closer to
the results of the DSSV fit. A noteworthy feature of the new global
fit is the observed pattern of the flavor symmetry breaking.
It indicates that $\Delta \bar{u}(x) > 0$ and $\Delta \bar{d}(x) < 0$
with the magnitude correlation
$|\Delta \bar{u}(x)| < |\Delta \bar{d}(x)|$.
We emphasize that this characteristic of the flavor symmetry
breaking pattern of the polarized sea-quark distribution is just
what the CQSM predicts \cite{Waka03AB}.
An interesting question is therefore how
this unique pattern of symmetry violation arises in the CQSM.
Since the physics is basically common in two versions of the CQSM,
we explain it in simpler SU(2) CQSM.
To this end, we first recall the fact that, within the theoretical
framework of the CQSM, the isoscalar and isovector distributions have
different theoretical structure due to their different
$N_c$-dependence \cite{DPPPW9697},\cite{WK99},\cite{Waka03AB},
so that the longitudinally polarized distribution
functions with each flavor is evaluated as linear combinations of
the isoscalar and isovector parts as
\begin{eqnarray}
 \Delta \bar{u}(x) &=& \frac{1}{2} \,\left[\,
 (\Delta \bar{u}(x) + \Delta \bar{d}(x)) \ + \ 
 (\Delta \bar{u}(x) - \Delta \bar{d}(x)) \,\right] , \\
 \Delta \bar{d}(x) &=& \frac{1}{2} \,\left[\,
 (\Delta \bar{u}(x) + \Delta \bar{d}(x)) \ - \ 
 (\Delta \bar{u}(x) - \Delta \bar{d}(x)) \,\right]. 
\end{eqnarray}

\begin{figure}[htb]
\begin{center}
  \includegraphics[height=.35\textheight]{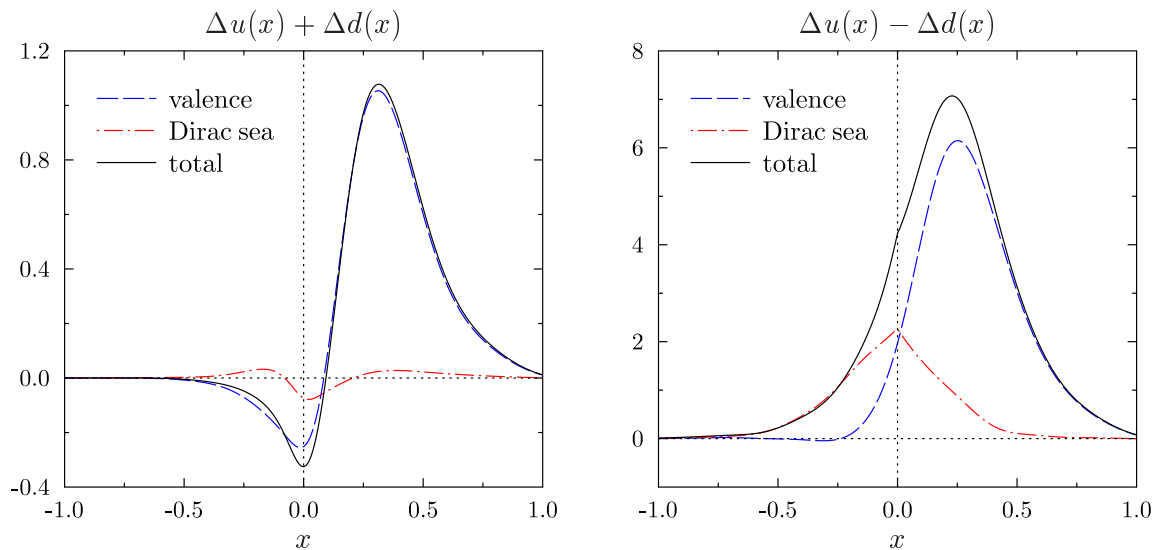}
  \caption{\baselineskip16pt The predictions of the SU(2) CQSM for
  the isoscalar and isovector longitudinally polarized distribution
  functions of the nucleon at the scale of the model.
  The long-dashed and the dash-dotted curves respectively stand for
  the contribution of $N_c$ valence quarks and that of the Dirac-sea
  quarks, while the solid curve represents their sum.
  The distribution functions $\Delta q(x)$ in the negative $x$ region
  should be interpreted as antiquark distributions according to the
  rule : $\Delta \bar{q}(x) = \Delta q(-x)$ with $0 < x < 1$.}
\label{fig3:lgpolmodel}
\end{center}
\end{figure}

Shown in Fig.3 are the predictions of the SU(2) CQSM for the
isoscalar and isovector combinations of the longitudinally polarized
quark distribution functions \cite{WK99}.
In this figure, the distribution functions with negative value of
$x$ should be interpreted as antiquark distribution according
to the rule : 
\begin{eqnarray}
 \Delta \bar{u}(x) \pm \Delta \bar{d}(x) &=& 
 \Delta u(-x) \pm \Delta d(-x) \hspace{10mm} (0 < x < 1).
\end{eqnarray}
The right panel of Fig.3 shows that the vacuum polarization of the
Dirac-sea quarks in the hedgehog mean-field plays an important role
in generating large flavor asymmetry of the polarized sea-quark
distribution. This fact was already emphasized in several previous
papers \cite{WW00},\cite{DGPW00}.
The physics we are now discussing is connected with another
unique feature of the CQSM predictions. As shown in the left panel of
Fig.3, it predicts that the isoscalar longitudinally
polarized distribution is negative in the small $x$ region including
the negative $x$ domain \cite{WK99},\cite{Waka03AB}, which means that
$\Delta \bar{u}(x) + \Delta \bar{d}(x)$ is negative for the physical
value of $x$, i.e. for $0 < x < 1$. This observation, combined with
the fact that $\Delta \bar{u}(x) - \Delta \bar{d}(x)$ is sizably large
and positive for small and negative $x$, leads to an interesting
symmetry breaking pattern of the
longitudinally polarized sea-quark distributions such that
$\Delta \bar{u}(x) > 0$ and $\Delta \bar{d}(x) < 0$ with the
magnitude correlation $|\Delta \bar{u}(x)| < |\Delta \bar{d}(x)|$.
We emphasize that this feature comes about as a parameter-free
prediction of the CQSM.
Some years ago, we have pointed out \cite{Waka07} that the negativity
of the isoscalar longitudinally polarized quark distribution in the
small $x$ region is just what is required for reproducing the sign
change of the deuteron spin structure function at low $x$
as dictated by the SMC and COMPASS data \cite{COMPASS05},\cite{SMC98}.
(Remember that the deuteron spin structure function is roughly
proportional to the isoscalar longitudinally polarized distribution
function of the nucleon.) 
This observation then indicates that the flavor
symmetry breaking pattern $|\Delta \bar{u}(x)| < |\Delta \bar{d}(x)|$
obtained in the DSSV fit must be strongly influenced by the deuteron
structure function data included in their global fit.

\begin{figure}[htb]
\begin{center}
  \includegraphics[height=.4\textheight]{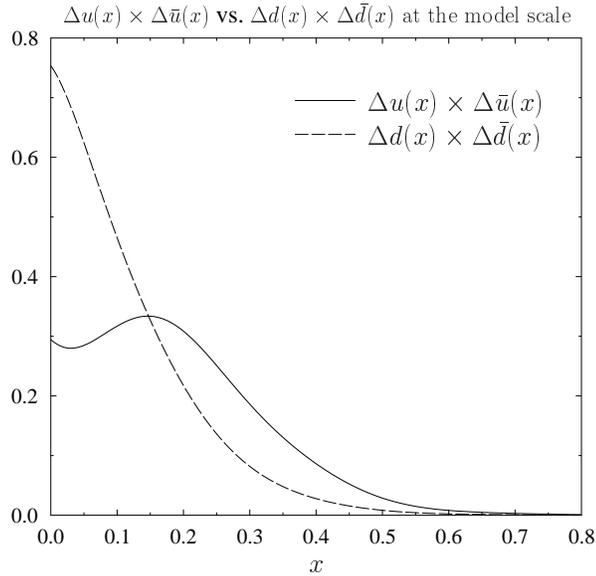}
  \caption{\baselineskip16pt The predictions of the SU(2) CQSM
  for the product of the polarized quark and antiquark
  distributions, i.e. $\Delta u(x) \times \Delta \bar{u}(x)$ and
  $\Delta d(x) \times \Delta \bar{d}(x)$ at the model scale.
  The expectation from the Pauli exclusion principle argument is
  an approximate equality of these two quantities, i.e.
  $\Delta u(x) \times \Delta \bar{u}(x) \simeq
  \Delta d(x) \times \Delta \bar{d}(x)$.}
\label{fig4:quauqdad}
\end{center}
\end{figure}

We recall that the flavor asymmetry of the polarized sea-quark
distributions is predicted also by some models with more
phenomenological nature like the statistical parton
model \cite{Bhalerao01},\cite{BSB02} as well
as the model based on the Pauli exclusion principle \cite{GR00}.
For instance, the analysis by Bhalerao within the statistical
model predicts $\Delta \bar{u}(x) > 0$ and $\Delta \bar{d}(x) < 0$
with the magnitude of $|\Delta \bar{d}(x)|$ being larger
than that of $|\Delta \bar{u}(x)|$ although slightly \cite{Bhalerao01}.
Note, however, that this model
uses the known empirical information for the magnitudes of
$\Delta u + \Delta \bar{u}$, $\Delta d + \Delta \bar{d}$, and
$\Delta s + \Delta \bar{s}$, so that its predictions are not of
purely theoretical nature. Also interesting is the prediction of
the semi-phenomenological model of G\"{u}ck and Reya based on the
Pauli exclusion principle \cite{GR00}.
Their model also predicts large flavor
asymmetry for both of the unpolarized and polarized sea-quark
distributions.
Furthermore, this model predicts fairly large asymmetry for the
magnitudes of $\Delta \bar{u}(x)$ and $\Delta \bar{d}(x)$ such that
$|\Delta \bar{u}(x)| < |\Delta \bar{d}(x)|$ in conformity with the
new DSSV global fit. We point out that the above-mentioned feature
comes from the basic ansatz of this semi-phenomenological treatment,
which demands that the product
\begin{equation}
 \Delta q(x,\mu^2) \,\Delta \bar{q}(x,\mu^2) \ \equiv \ P(x),
\end{equation}
is universal flavor-independent function $P(x)$ with $\mu$ being
an low energy input scale of their evolution program, since
the effect of Pauli blocking is only related to the spin of quarks
and antiquarks irrespective of their flavor degrees of freedom.
Since it is empirically known that $|\Delta u(x)| > |\Delta d(x)|$,
it naturally follows that $|\Delta \bar{u}(x)| < |\Delta \bar{d}(x)|$.
It may be of some interest to check to what extent the above ansatz
holds in our explicit dynamical model predictions. Shown in Fig.4 are
the predictions of the SU(2) CQSM for the product of $\Delta u(x)$ and
$\Delta \bar{u}(x)$ and that of $\Delta d(x)$ and $\Delta \bar{d}(x)$
at the model energy scale, which we identify with
$\mu^2 = 0.30 \,\mbox{GeV}^2$. One clearly sees that the ansatz
$\Delta u(x) \,\Delta \bar{u}(x) = \Delta d(x) \,\Delta \bar{d}(x)$
does not hold good at least in the CQSM. This seems to be an
indication that the nontrivial chiral dynamics of QCD besides
the Pauli blocking effect plays some important roles in the
physics of the flavor asymmetry of sea-quark distributions in the
nucleon.

\begin{figure}[htb]
\begin{center}
  \includegraphics[height=.4\textheight]{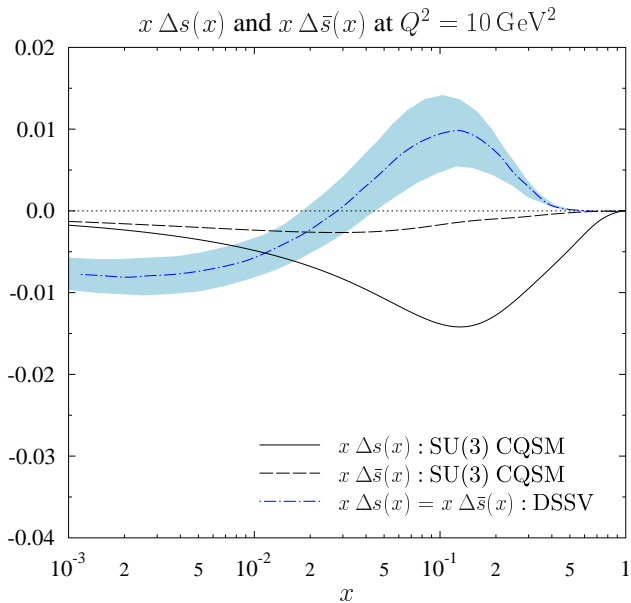}
  \caption{\baselineskip16pt The prediction of the $SU(3)$ CQSM
  for the polarized strange quark distribution
  $x \,\Delta s(x)$ in comparison with the DSSV global fit.
  Here, the solid curve is the prediction of the SU(3) CQSM for
  the polarized $s$-quark distribution, whereas the long-dashed
  curve is that for the polarized $\bar{s}$-quark distribution.} 
\label{fig4:xDelsbar}
\end{center}
\end{figure}

Finally, we make a brief comment on the polarized strange quark
distributions obtained in the DSSV global analysis.
As shown in Fig.5, a remarkable feature of the new DSSV analysis
is that a polarized strange quark distribution $\Delta s(x)$ is
positive at large or medium $x$, but negative at small $x$, at
variance with most of the past PDF fits which use only inclusive
DIS (deep-inelastic-scattering) data.
This peculiar behavior of $\Delta s(x)$ arises since the (kaon)
semi-inclusive DIS data prefer a small and likely positive
$\Delta s(x)$ at medium $x$, while inclusive DIS and the
constraints from beta-decay demand a negative 1st moment of
$\Delta s(x)$ \cite{LS03}, thereby forcing $\Delta s(x)$ to be
negative at small $x$.
To our knowledge, there is no theoretical model, which predicts
such nodal behavior of $\Delta s(x)$. Shown in Fig.5 together with
the DSSV fit are the predictions of the SU(3) CQSM for the polarized
strange quark distributions. 
The SU(3) CQSM predicts that both
of $\Delta s(x)$ and $\Delta \bar{s}(x)$ are negative in the whole
region of $x$, while the magnitude of $\Delta \bar{s}(x)$ is
much smaller than $\Delta s(x)$, i.e.
$|\Delta \bar{s}(x)| \ll |\Delta s(x)|$.
In the DSSV analysis, the equality of
the polarized strange quark and antiquark distributions is
assumed, because of the reason that the fit is unable to
discriminate strange quarks from
antiquarks at the present stage.  
As pointed out by the authors of \cite{DSSV09} themselves, however,
unlike the spin-averaged case where the distributions of $s(x)$ and
$\bar{s}(x)$ are constrained by the conservation law, i.e.
$\int_0^1 \,[s(x) - \bar{s}(x)] \,dx = 0$, there is no absolute
need for $\Delta s(x)$ and $\Delta \bar{s}(x)$ to have the same
magnitude or even the same sign.  
Although may not be feasible at the present stage,
a possible large asymmetry of the polarized strange
sea as suggested by the CQSM should be kept in mind and
such possibility is highly desirable to be taken into account
in more elaborate global analyses in the future.

To sum up, the recent global analysis of spin-dependent parton
distributions appears to offer the first strong evidence in favor
of the flavor symmetry violation of the longitudinally polarized
sea-quark distributions in the nucleon.
We have pointed out that the indicated flavor symmetry breaking
pattern, i.e. $\Delta \bar{u}(x) > 0$ and $\Delta \bar{d}(x) < 0$
with the magnitude correlation
$|\Delta \bar{u}(x)| < |\Delta \bar{d}(x)|$, is remarkably
consistent with the nearly-parameter-free prediction of the CQSM.
An apparent discrepancy remains, however, between their fit for
the polarized strange quark distributions in the nucleon and
the corresponding prediction of the SU(3) CQSM. To get more
definite conclusion on the implication of this discrepancy, we
certainly need more and more effort to understand the precise
mechanism of semi-inclusive reactions, especially the mechanism
of semi-inclusive kaon productions.

\vspace{16mm}
\noindent
\begin{large}
{\bf Acknowledgment}
\end{large}

\vspace{3mm}
This work is supported in part by a Grant-in-Aid for Scientific
Research for Ministry of Education, Culture, Sports, Science
and Technology, Japan (No.~C-A215402680)

%
%

\setlength{\baselineskip}{5mm}


\begin{thebibliography}{99}

\bibitem{NMC91}
NMC Collaboration, P.~Amaudruz~et al.,
Phys. Rev. Lett. 66 (1991) 2712.

\bibitem{Kumano98}
S.~Kumano, Phys. Rep. 303 (1998) 183.

\bibitem{GP01}
G.T.~Garvey and J.C.~Peng, Prog. Part. Nucl. Phys. 47
(2001) 203.


\bibitem{Sullivan72}
J.D.~Sullivan, Phys. Rev. D5 (1972) 1732.

\bibitem{Thomas83}
A. W. Thomas, Phys. Lett. B126 (1983) 97.

\bibitem{HM90}
E.M.~Henley and G.A.~Miller, Phys. Lett. B251 (1990) 453.

\bibitem{Burk02}
S.~Kumano, Phys. Rev. D43 (1991) 3067.

\bibitem{MT93}
W.~Melnitchouk and A.W.~Thomas, Phys. Rev. D47 (1993) 3794.

\bibitem{HSS96}
H.~Holtman, A.~Szczurek, and J.~Speth,
Nucl. Phys. A596 (1996) 631.

\bibitem{Waka91}
M.~Wakamatsu, Phys. Rev. D44 (1991) R2631.

\bibitem{Waka92}
M.~Wakamatsu, Phys. Rev. D46 (1992) 3762.

\bibitem{WK98}
M.~Wakamatsu and T.~Kubota, Phys. Rev. D57 (1998) 5755.

\bibitem{PPGWW99}
P.V.~Pobylitsa, M.V.~Polyakov, K.~Goeke, T.~Watabe, and C.~Weiss,\\
Phys. Rev. D59 (1999) 034024.

\bibitem{Bhalerao01}
R.S.~Bhalerao, Phys. Rev. C63 (2001) 025208.

\bibitem{BSB02}
C.~Bourrely, J.~Soffer, and F.~Buccella, Eur. Phys. J. C23 (2002) 487.

\bibitem{GR00}
M.~Gl\"{u}ck and E.~Reya, Mod. Phys. Lett. A15 (2000) 883.

\bibitem{FS98}
R.J.~Fries and A.~Sch\"{a}fer, Phys. Lett. B443 (1998) 40.

\bibitem{BK99}
K.G.~Boreskov, A.B.~Kaidalov, Eur. Phys. J. C10 (1999) 143.

\bibitem{CS01}
F.G.~Cao and A.I.~Signal, Eur. Phys. J. C21 (2001) 105.

\bibitem{KM02}
S.~Kumano and M.~Miyama, Phys. Rev. D65 (2002) 034012.

\bibitem{FSW03}
R.J.~Fries, A.~Sch\"{a}fer, and C.~Weiss, Eur. Phys. J. A17 (2003) 509.

\bibitem{CS03}
F.G.~Cao and A.I.~Signal, Phys. Rev. D68 (2003) 074002.

\bibitem{DPPPW9697}
D.I.~Diakonov, V.Yu.~Petrov, P.V.~Pobylitsa, M.V.~Polyakov, and
C.~Weiss, \\
Nucl. Phys. B480 (1996) 341 ; Phys. Rev. D56 (1997) 4069.
 
\bibitem{WGR9697}
H.~Weigel, L.~Gamberg, and H.~Reinhardt, Mod. Phys. Lett. 
A11 (1996) 3021 ; \\
Phys. Lett. B399 (1997) 287.

\bibitem{WK99}
M.~Wakamatsu and T.~Kubota, Phys. Rev. D60 (1999) 034020.

\bibitem{WW00}
M.~Wakamatsu and T.~Watabe, Phys. Rev. D62 (2000) 017506.

\bibitem{DGPW00}
B.~Dressler, K.~Goeke, M.V.~Polyakov, and C.~Weiss,
Eur. Phys. J. C14 (2000) 147. 

\bibitem{Waka03AB}
M.~Wakamatsu, Phys. Rev. D67 (2003) 034005 ; 
Phys. Rev. D67 (2003) 034006.

\bibitem{WY91}
M.~Wakamatsu and H.~Yoshiki, Nucl. Phys. A524 (1991) 561.

\bibitem{WW00L}
M.~Wakamatsu and T.~Watabe, Phys. Rev. D62 (2000) 054009.

\bibitem{Waka09}
M.~Wakamatsu, arXiv:0908.0972 [hep-ph].

\bibitem{Thomas09}
A.W.~Thomas, Phys. Rev. Lett. 101 (2009) 102003.

\bibitem{WT05}
M.~Wakamatsu and H.~Tsujimoto, Phys. Rev. D71 (2005) 074001.

\bibitem{WN0608}
M.~Wakamatsu and Y.~Nakakoji, Phys. Rev. D74 (2006) 054006 ; \\
Phys. Rev. D77 (2008) 074011.

\bibitem{HERMES0504}
HERMES Collaboration : A.~Airapetian et al., 
Phys. Rev. D71 (2005) 012003 ; \\
Phys. Rev. Lett. 92 (2004) 012005.

\bibitem{DSSV09}
D.~de~Florian, R.~Sassot, M.~Stratmann, and W.~Vogelsang,
Phys. Rev. D80 (2009) 034030.

\bibitem{DPP88}
D.I.~Diakonov, V.Yu.~Petrov, and. P.V.~Pobylitsa, Nucl. Phys.
B306 (1988) 809.

\bibitem{WAR92}
H.~Weigel, R.Alkofer, and H.~Reinhardt, Nucl. Phys. B387 (1992) 638.

\bibitem{BDGPPP93}
A.~Blotz, D.I.~Diakonov, K.Goeke, N.W.~Park, V.Yu.~Petrov, and
P.V.~Pobylitsa, \\
Nucl. Phys. A555 (1993) 765. 

\bibitem{GRV95}
M.~Gl\"{u}ck, E.~Reya, and A.~Vogt, Z. Phys. C67 (1995) 433.

\bibitem{GRSV96}
M.~Gl\"{u}ck, E.~Reya, M.~Stratmann, and A.~Vogt, 
Phys. Rev. D53 (1996) 4775.

\bibitem{Waka07}
M.~Wakamatsu, Phys. Lett. B646 (2007) 24.

\bibitem{COMPASS05}
COMPASS Collaboration, E.S.~Ageev~et al.,
Phys. Lett. B612 (2005) 154.

\bibitem{SMC98}
SMC Collaboration, B.~Adeva~et al.,
Phys. Rev. D58 (1998) 112001.

\bibitem{LS03}
E.~Leader and D.B.~Stamenov, Phys. Rev. D67 (2003) 037503.

\end{thebibliography}
\end{document}